# Laser Induced Forward Transfer of conducting polymers

M. Kandyla[1], S. Chatzandroulis[2], and I. Zergioti*[1,a]

[1]*Physics Department, National Technical University of Athens, 9 Heroon Polytechniou Street, Zografou 15780, Athens, Greece*

[2] *Institute of Microelectronics, NCSR Demokritos, Agia Paraksevi 15310, Athens, Greece*

**Abstract:** We report on laser printing of conducting polymers directly from the solid phase. Laser Induced Forward Transfer is employed to deposit P3HT:PCBM films on glass/ITO/PEDOT:PSS substrates. P3HT:PCBM is widely used as the active material in organic solar cells. Polyaniline films, which are also printed by Laser Induced Forward Transfer, find many applications in the field of biotechnology. Laser printing parameters are optimized and results are presented. To apply solid-phase laser printing, P3HT:PCBM films are spun cast on quartz substrates, while aniline is in-situ polymerized on quartz substrates.

## 1. Introduction

Conducting polymers offer an attractive alternative to their inorganic counterparts due to properties such as flexibility, low cost, light weight, and ease of processing. Applications based on organic conductors and semiconductors include chemical and biological sensors [1,2], organic light emitting diodes and transistors

---

[a] Physics department, 9 Heroon Polytechniou st., Zografou 15780, Greece, email: zergioti@central.ntua.gr

[3], organic photovoltaics [4], conducting adhesives [5], batteries [6], printed circuit boards [7], and antistatic coatings [8].

Deposition and patterning of polymer materials has been achieved by a variety of methods, such as electropolymerization [2], spin-coating [4], inkjet printing [9], screen printing [10], and spraying deposition [11], among others. The aforementioned techniques require liquid-phase processing, therefore solvent properties become critical. Furthermore, liquid deposition complicates the fabrication of all-organic, multilayer structures due to the interaction between solvents and materials from different layers. In this paper, we present solid-phase deposition of polymer materials of increased technological interest, employing laser printing techniques. Laser printing is a non-contact deposition method, offering high spatial resolution and compatibility with sensitive materials. This paper is organized as follows. Section 2 outlines the technological value of organic solar cells, summarizes the most important advances in the field of bulk heterojunction photovoltaics, and presents results on solid-phase laser printing of active materials for solar cell applications. Section 3 describes the use of polyaniline in the field of biotechnology, addresses the challenges of fabricating microsystems based on this material, and presents results on laser printing of polyaniline films for biotechnological applications. Section 4 includes the discussion and conclusions based on the results presented in Sections 2 and 3.

**2. Laser printing for organic solar cells**

**2.1 Introduction**

Solar cell research is becoming increasingly important as the photovoltaic power industry experiences intense growth globally. Photovoltaics and wind power are forecast as the growth leaders among alternative electricity sources for the next 20 years. The advantages of photovoltaic power include lack of dependence on fuel, lack of carbon emissions, and lack of water use. So far the main disadvantage of photovoltaic devices has been the increased production cost, compared to other energy resources. However, the manufacturing cost of solar cells is continuously decreasing and the industry is expected to become competitive within the next 5 years [12]. Therefore, photovoltaic research aiming at cost-effective technologies is becoming more relevant than ever. Thin-film solar cells offer the advantage of lower cost compared to crystalline silicon devices. Among them, polymer-based organic solar cells offer the additional advantages of light weight, mechanical flexibility, and ease of processing, while their properties can additionally be tuned by adjusting the chemical composition of the active layer.

One class of polymer photovoltaic cells showing increased potential in terms of power conversion efficiency and long-term stability is bulk heterojunction solar cells [4], with an active layer of a conjugated polymer blended with fullerene derivatives [13]. These two materials form an interpenetrating bicontinuous network with large interfacial area, allowing for efficient exciton dissociation and charge transport upon illumination. The most commonly employed technique for the fabrication of bulk heterojunction solar cells is spin-coating. Alternatively, non-contact printing techniques such as inkjet printing and airbrush spray deposition, among others, have also been used. Printing technologies allow for high volume and low cost processing. Additionally, printing offers the possibility of selectively depositing the active layer on specific areas of the substrate as opposed to spin-coating. However, both spin-coating and printing require the active material to be deposited in the form of a solution. This requirement hinders the realization of all-organic solar cells, in which the electrodes as well as the active layer are composed of organic materials, because the solvent of each layer may interact with the layers lying underneath.

Spin-coating deposition of bulk heterojunction solar cells involves the active material being spun cast in the form of a solution on a transparent, electrically conducting substrate (Fig. 1). Usually the substrate consists of a glass surface covered by indium tin oxide (ITO), on which a layer of poly(3,4-ethylenedioxythiophene):polystryrene sulfonate (PEDOT:PSS) is spun cast. ITO is a transparent conducting material, which serves as the anode electrode of the device, while allowing for illumination of the active layer. PEDOT:PSS is also a transparent conducting polymer, which helps decrease the hole barrier for charge collection. In certain cases, PEDOT:PSS modified by the addition of a polar compound, which increases the conductivity of PEDOT, has been used as the transparent anode electrode instead of ITO [14]. The active layer, which is typically composed of a solution of poly(3-hexylthiophene) (P3HT) and [6,6]-phenyl-C61-butyric acid methyl ester (PCBM), is then spun cast on the substrate. PCBM is an electron-conducting fullerene derivative, which acts as the acceptor material, while P3HT is a hole-conducting polymer, which acts as the donor material. The most frequently used solvents for the P3HT:PCBM blend are chlorobenzene [9,15], dichlorobenzene [14,16], chloroform [17,18], and trichlorobenzene [19,20]. Often fabrication of the device is done in a glove box [15,21] in order to avoid oxidation effects, although fabrication in air has been reported as well [9,16]. The cathode electrode is then deposited through a shadow mask, which defines the active area of the device. The cathode electrode is typically a metal, such as Al [15,19], Au [16], or other low work function metals [22]. Power conversion efficiencies up to 5-6% have been achieved with spin-coated P3HT:PCBM solar cells [23,24,25,26], which are among the highest efficiency values for organic solar cells so far.

Inkjet printing has also been used for the deposition of the active layer in bulk heterojunction solar cells. Inkjet printing is a non-contact technique, in which the printhead does not come into direct contact with the substrate, thus presenting certain advantages compared to other printing methods. Often, a high boiling-point solvent [9] is used in addition to the common P3HT:PCBM solvents in order to avoid clogging of the printhead, which is a typical problem in inkjet printing. The quality of the ejected drop is of critical importance and depends on a number of factors, such as the choice of solvents, the solvent mixture ratio, and the weight percentage of P3HT:PCBM in the solution. Great care is taken to optimize these factors for printing organic solar cells [9]. Droplets and continuous films deposited by inkjet printing suffer from the coffee-drop effect [27], due to which solute material accumulates at the edge. The regioregularity of the polymer donor is also an important parameter. Even though spin-coated organic solar cells show higher efficiencies for higher values of regioregularity of the polymer donor, high regioregularity induces rapid gelling time during inkjet printing, which sets an upper limit on the regioregularity of the polymers used by this method [28]. The resolution of inkjet printing is set by the minimum dimension which can be deposited by this technique. The diameter of printed drops on the substrate is significantly bigger than the diameter of the printhead nozzle, due to spreading of the jetted volume. Inkjet-printed droplets present minimum diameters on the order of 120 μm [9]. Power conversion efficiencies up to 3.5% have been achieved by inkjet printing of organic solar cells [28].

Another non-contact printing technique for the fabrication of bulk heterojunction solar cells is airbrush spray deposition. This technique also uses a solution of the active material, which is sprayed on the device substrate by a handheld airbrush. However, because many coating layers are needed to achieve complete coverage of the substrate, the films produced by airbrush spraying suffer from inhomogeneities, which affect the film quality [29]. The performance of organic solar cells prepared by airbrush spray deposition depends heavily on the solvent used. It turns out that chlorobenzene is the most suitable solvent for this technique, while dichlorobenzene and trichlorobenzene, which are commonly used for spin-coating of organic solar cells, are not well-suited for airbrush spraying [29]. Power conversion efficiencies of 2.3-2.8% have been achieved by airbrush spray deposition [11,29].

Several methods have been employed in order to increase the power conversion efficiency of organic solar cells. These methods include phase separation of P3HT and PCBM in the active layer blend [19], reduction of the solvent evaporation speed by using high boiling-point solvents [30] and solvent-vapour treatment [20,31], electrode tailoring [16,32], patterning and doping of the PEDOT:PSS layer [25,33], thermal annealing [18,34,35], modified device architecture [21], treatment with an applied external potential [36], and microwave irradiation [37], among others. Even though the power conversion efficiency of organic solar cells is still below that of inorganic devices, their remarkable mechanical and optical properties make them ideal for applications in consumer electronics and smart fabrics [38].

## 2.2 Experimental

In this work, Laser Induced Forward Transfer (LIFT) [39] was used for the deposition of the active material for solar cell applications. LIFT is a non-contact printing technique, which is based on the controlled transfer of a thin film, from a transparent carrier to a receiving substrate, using one or more laser pulses (Fig. 2). The material to be transferred, also called the donor material, is deposited on one side of a transparent carrier, usually made of glass or quartz depending on the wavelength of the laser system. The receiving substrate is then placed parallel to the carrier, facing the donor material. The distance between the carrier and the substrate is typically very short, ranging from near contact to a few μm. A laser pulse incident on the back side of the transparent carrier is focused on the donor material and, by means of shock formation and ablation, ejects a small part of the donor film forward, resulting in the deposition of the ejected material on the substrate. The size of the transferred material depends on the size of the focused laser beam that irradiates the donor film surface. By varying the focusing conditions of the beam we can control the size of the printed material. Additionally, by placing a mask on the laser beam path and imaging the mask profile on the carrier/donor material interface, we can further vary the shape and size of the printed material. We can repeat this process by translating the carrier between successive laser pulses in order to irradiate a fresh spot on the carrier/donor film interface each time. Furthermore, we can choose to translate the substrate as well, in order to print two-dimensional patterns of the donor material. Laser printing has already been used successfully for the fabrication of organic electronic devices, such as organic thin-film transistors [40] and light-emitting diodes [41,42].

P3HT:PCBM films were LIFT printed onto various substrates for photovoltaic applications. A 1:1 w/w solution of 15 mg/ml P3HT (Rieke Metals, Inc.) and 15 mg/ml PCBM (Nano-C, Inc.) is prepared in chloroform. The solution is spun cast on a quartz target, which acts as the transparent carrier for LIFT deposition. We choose quartz and not plain glass as the carrier because only quartz is transparent at the wavelength of the laser beam (266 nm) that irradiates the carrier/donor interface. Chloroform evaporates after spin-coating and a solid film of P3HT:PCBM forms on the quartz carrier. We use this film as the donor for solid-phase LIFT deposition.

In this work, plain glass and glass/ITO/PEDOT:PSS were used as substrates. Plain glass substrates are cleaned with a ten minute ultrasonic bath in acetone, followed by a ten minute ultrasonic bath in methanol. The substrates are then rinsed with double deionized water and dried with a nitrogen gas flow. Glass/ITO substrates (VisionTek Systems, sheet resistance 15 Ω/ ) are cleaned with a sequence of ultrasonic baths, first five minutes in acetone, then five minutes in isopropanol, then five minutes in double deionized water, and finally five minutes in acetone. The substrates are dried with a nitrogen gas flow and baked at 100 °C for ten minutes. A thin layer of PEDOT:PSS (Clevios P VP.AI 4083) is then spun cast on the glass/ITO substrates at 1800 rpm for 20 s and the substrates are baked at 120 °C for ten minutes.

The experimental apparatus is shown in Fig. 2. The laser beam, consisting of the 4$^{th}$ harmonic of a Q-switched Nd:YAG laser (4 ns pulse duration, 266 nm wavelength), is propagated through a variable rectangular mask, which selects the middle part of the beam and reduces its size. The selected part of the beam is then imaged onto the quartz/P3HT:PCBM interface, through a system of lenses and a 15x microscope objective. The substrate is placed in close proximity to the donor carrier, at a distance less than 50 μm. A single laser pulse irradiates the P3HT:PCBM surface and induces material transfer on the underlying substrate.

**2.3 Results**

Figure 3 shows an optical microscope image of P3HT:PCBM spots, deposited on a glass substrate by the LIFT method. The deposition was performed entirely in the solid phase. The rectangular mask through which the laser beam is propagated (Fig. 2), is imaged on the P3HT:PCBM donor film and determines the shape and size of the printed spots. Figure 3 shows printing results for different imaging conditions of the rectangular mask, where the top row is closer to the mask being perfectly imaged on the donor film, while in the rows that follow the mask image is gradually becoming defocused. We observe how the effect of defocusing affects the shape of the spots that appear in Fig. 3, where at the top row the spots have sharper edges, in contrast to the bottom row where the spots are larger with less well-defined edges. Apart from the imaging conditions, the laser fluence was varied as well. The results shown in Fig. 3 indicate that laser printing of P3HT:PCBM films can be achieved for a wide range of laser fluences, therefore the technique is resilient to pulse-to-pulse intensity fluctuations. For fluences significantly above 170 mJ/cm$^2$ the quality of the deposited spots deteriorates and black dots appear consistently on the surface, indicating the material is being damaged by the laser pulse. The deposited spots are between 100-200 μm wide. This dimension can be adjusted in a controlled manner by adjusting the dimensions of the rectangular mask. Using a profilometer, we find the thickness of the spots to be 190 nm. Most often, P3HT:PCBM film thickness for photovoltaic applications ranges between 100-250 nm [31,36]. This has been found to be an optimal range, given there is an interplay between low charge mobility in organic materials that requires thinner active layers for efficient charge collection, and increased light absorption that requires thicker active layers. Finally, the adhesion of the spots on the glass substrate is very good, as they remain in place after the deposition until today. Printing of P3HT:PCBM on glass/ITO/PEDOT:PSS substrates produced similar results as the ones presented in Fig. 3.

The lateral dimensions of P3HT:PCBM films do not exceed 200 μm in the results presented here. For devices with active areas in the range of cm$^2$, we can print several adjacent spots to create a larger active surface, while simultaneously increasing the rectangular mask opening, which results in an increase in the individual spot dimensions. Adjacent deposition can happen in an automated way and create solar cells of any two-dimensional pattern possible.

## 3. Laser printing of polyaniline films

### 3.1 Introduction

Polyaniline is one of the most commonly used conducting polymers due to its high conductivity, excellent environmental stability, and ease of preparation. In the field of biotechnology, polyaniline is employed in biological sensors [43], biomembranes [44], and as a substrate for mammalian cell growth [45]. The practical use of polyaniline is hindered by its mechanical properties and its poor solubility in most common organic solvents. In order to overcome this difficulty, polyaniline is usually prepared in the form of dispersions suited to conventional application methods [46]. In most biosensing applications polyaniline is obtained through electrochemical polymerization on the surface of the working electrode [47]. Often entrapment of biological molecules during the electropolymerization process is employed, in order to render the polyaniline substrates active for biosensing applications [48]. However, the need for processing biomaterials within a neutral pH range leads to electro-inactivity of the deposited films, discouraging the use of polyaniline as a biosensing material [49]. Inkjet printing is an alternative method of depositing conducting polymers for the fabrication of biosensors [46,50]. Nevertheless, most inkjet-printing applications require sophisticated synthetic procedures in order to avoid particle–particle aggregation in the dispersion medium. Therefore, the development of a simple and reliable method for fabricating polyaniline microsensors is still a challenge.

We now present results on the deposition of polyaniline films directly from the solid phase, without the use of solvents or electrochemistry, by using the LIFT method. We are able to control the dimensions of the deposited films and achieve spatial resolution of a few μm, thus enabling the use of polyaniline in microsystems. Polyaniline films are formed in-situ on transparent quartz carriers and a nanosecond laser is used to transfer selected areas of controllable dimensions on a glass substrate.

## 3.2 Experimental

Anilinium chloride and ammonium peroxodisulphate were purchased from Panreac, Spain and used as received. For the synthesis of polyaniline films we followed the in-situ method described by Stejskal *et al*. [51]. 0.2 M anilinium chloride is oxidized with 0.2 M ammonium peroxodisulfate in 1.0 M HCl at room temperature (~20 °C). Quartz substrates are covered by adhesive tape on one side and placed in the reaction vessel, where they stay overnight. After the polymerization is over, the substrates are well rinsed with dilute HCl to remove the adhering polyaniline precipitate, and dried. The quartz samples covered by a thin polyaniline film are then separated from the adhesive tape and used as carriers for LIFT.

We performed LIFT of polyaniline on glass substrates, using the experimental apparatus described in Section 2.2. By focusing one laser pulse on the polyaniline surface, we are able to deposit square spots of polyaniline on the substrate.

### 3.3 Results

Figure 4 shows an optical microscope image of polyaniline spots deposited by LIFT on a glass substrate for various incident laser fluences and various imaging configurations of the rectangular mask, through which the laser beam is propagated before irradiating the polyaniline donor surface. All spots were deposited using one laser pulse. Changes in spot morphology from row to row are mainly due to slight variations in the donor film morphology across the surface of the quartz carrier, from which the spots originate. Using a profilometer, we find the average thickness of the spots to be 180 nm. Diffraction effects across the boundaries of the rectangular spots become visible in the last two rows, which correspond to the most defocused imaging of the mask. Similar diffraction effects have been reported elsewhere [52]. Given that the fluence changes significantly between the first and the third column, it is evident that LIFT of polyaniline films can be applied for a conveniently wide range of laser parameters. For fluences significantly lower than the fluence range depicted in Fig. 4 there was no deposition of polyaniline films, while for fluences significantly higher than the fluence range in Fig. 4 the transferred films were destroyed by intense laser irradiation.

Laser deposition overcomes many of the disadvantages of polyaniline. Its poor mechanical properties and lack of solubility in organic solvents do not affect the LIFT method. Additionally, even though the in-situ polymerization used in this work takes place in the presence of excess HCl in an aqueous solution, the polymer film is subsequently removed from the solution and is meant to interact with biomaterials strictly in the solid phase, therefore the increased pH of the polymerizing solution does not affect the properties of sensitive biomolecules.

## 4. Discussion and Conclusions

The LIFT method offers several advantages compared to other printing techniques. It is a non-contact method and allows for the printing of sensitive materials on sensitive substrates. Compared to other non-contact printing techniques presented above, it is versatile and can be used with a variety of donor materials, including metals [53,54], semiconductors [55], organic materials [56,57] and sensitive biomaterials [58,59] such as DNA and proteins. It is compatible with any solvent in which organic donor materials may be dissolved. It does not suffer from printhead clogging, like inkjet printing, or from film inhomogeneity, like spray deposition. It can print two-dimensional patterns without the use of expensive lithographic masks and elaborate vacuum facilities. All the information about the shape of the pattern is stored digitally in the motion of the substrate. Finally, the biggest advantage of LIFT is that it is the only non-contact deposition technique with the ability to print the donor material both in the liquid and in the solid phase. One can choose to deposit a liquid layer of the donor material on the transparent carrier and instantly irradiate it with the laser beam so that a droplet is printed on the substrate or spin coat the surface of the carrier with the donor material, let it dry so that it forms a thin film, and then transfer a solid part of the donor film on the substrate. LIFT is the only non-contact deposition method that offers printing in the solid phase, which makes it essentially solvent-independent, since by the time printing occurs the solvent has evaporated. Therefore, one can select the best available solvent for a material to be printed, without being constrained by LIFT. Additionally, solid-phase printing is the most promising way to create all-organic devices, where one can print successive layers of organic materials, avoiding the interaction of the solvent of each layer with layers lying underneath. Finally, solid-phase LIFT presents fewer health risks since the operating personnel do not have to inhale toxic solvents during the deposition.

Solid-phase LIFT was employed for the deposition of P3HT:PCBM films on plain glass and glass/ITO/PEDOT:PSS substrates. Printing is achieved for a wide range of laser fluences. The adhesion of LIFT-deposited P3HT:PCBM spots on the substrate is very good. Even though the results presented here are for P3HT:PCBM films spun cast on quartz carriers from chloroform solutions, the printing process is solvent independent and donor films can be obtained for any P3HT:PCBM solvent. Two-dimensional patterns of various sizes can be created by translating the substrate between laser pulses.

Polyaniline films were successfully prepared by in-situ polymerization on quartz and used as donor films for solid-phase LIFT printing of polyaniline on glass substrates. By varying the dimensions of the mask through which the laser beam is propagated, we achieve printing of micrometer-size polyaniline spots, which can be used as substrates for microsystems. By varying the polymerization temperature, we can control the thickness of the polyaniline donor and eventually the thickness of the printed film.

**Acknowledgments**

We would like to thank Dr. C. Pandis for his help with polyaniline films. The research leading to these results was supported by the European Commission under a Marie Curie International Reintegration Grant, Seventh Framework Program, grant agreement 224790.

**Figure Captions**

**Figure 1**: Schematic representation of an organic solar cell.

**Figure 2**: Schematic representation of the Laser Induced Forward Transfer setup.

**Figure 3**: Optical microscope image of P3HT:PCBM spots deposited by Laser Induced Forward Transfer on a glass substrate for various laser fluences and various imaging conditions of the rectangular mask.

**Figure 4**: Optical microscope image of polyaniline spots deposited by Laser Induced Forward Transfer on a glass substrate for various laser fluences and various imaging conditions of the rectangular mask.

**Figures**

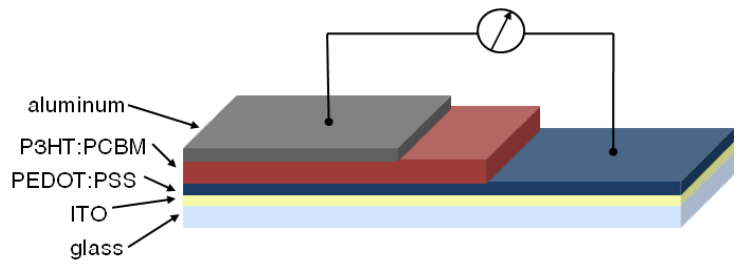

**Figure 1**

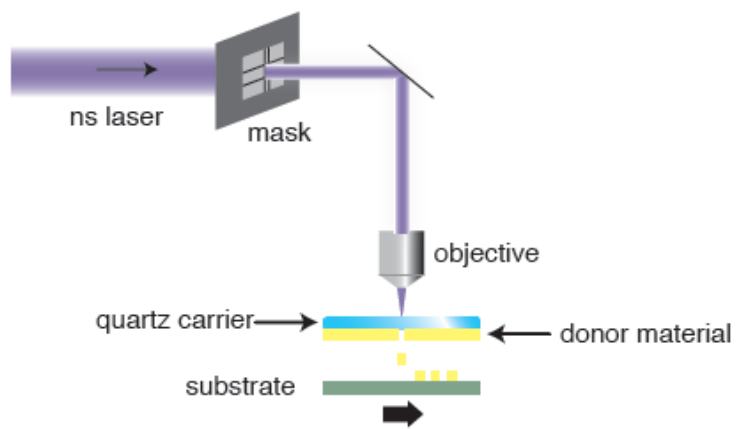

**Figure 2**

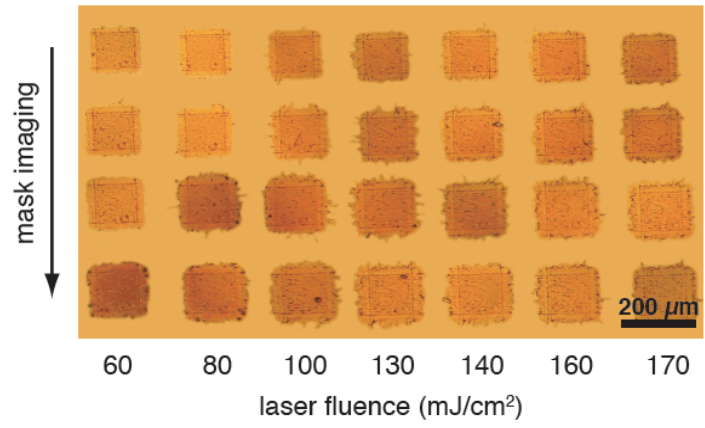

**Figure 3**

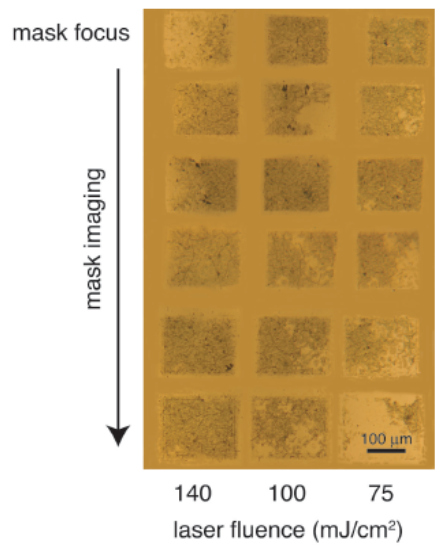

**Figure 4**